\documentclass[conference,table]{IEEEtran}
\IEEEoverridecommandlockouts

\usepackage{cite}
\usepackage{amsmath,amssymb,amsfonts}
\usepackage{algorithmic}
\usepackage{graphicx}
\usepackage{textcomp}
\usepackage[compact]{titlesec}
\usepackage{comment}

\usepackage{tikz}
\usetikzlibrary{trees}
\usetikzlibrary{positioning}
\usepackage{tikz-qtree}
\usetikzlibrary{trees}

\usepackage{hyperref}

\hypersetup{
    colorlinks=true,
    linkcolor=blue,
    filecolor=blue,      
    urlcolor=blue,
    citecolor=blue
}

\usepackage[utf8]{inputenc}
\usepackage{xcolor}
\usepackage{color, colortbl}
\usepackage[super]{nth}

\newcommand{\empcirc}[2][black,fill=white]{\tikz[baseline=-0.5ex]\draw[#1,radius=#2] (0,0) circle ;} 

\newcommand{\fillcirc}[2][black,fill=black]{\tikz[baseline=-0.5ex]\draw[#1,radius=#2] (0,0) circle ;}

\usepackage{fancyhdr}
\fancypagestyle{firstpage}{
  \fancyhf{}
  \fancyhead[r]{2020 \nth{23} International Conference on Computer and Information Technology (ICCIT), 19-21 December, 2020.}
  \fancyfoot[l]{978-0-7381-2333-2/20/\$31.00 ©2020 IEEE}
}

\def\BibTeX{{\rm B\kern-.05em{\sc i\kern-.025em b}\kern-.08em
    T\kern-.1667em\lower.7ex\hbox{E}\kern-.125emX}}
\begin{document}

\title{A Survey on Blockchain \& Cloud Integration}

\author{\IEEEauthorblockN{Soumik Sarker, Arnob Kumar Saha, Md Sadek Ferdous}
\IEEEauthorblockA{Department of Computer Science and Engineering, 
Shahjalal University of Science and Technology, Sylhet-3114, Bangladesh.\\}
\IEEEauthorblockA{Email: ronodhirsoumik@gmail.com, arnobkumarsaha00@gmail.com, sadek-cse@sust.edu\\}}

\maketitle
\thispagestyle{firstpage}

\begin{abstract}
Blockchain is one of the emerging technologies with the potential to disrupt many application domains. Cloud is an on-demand service paradigm facilitating the availability of shared resources for data storage and computation. In recent years, the integration of blockchain and cloud has received significant attention for ensuring efficiency, transparency, security and even for offering better cloud services in the form of novel service models. In order to exploit the full potential of blockchain-cloud integration, it is essential to have a clear understanding on the existing works within this domain. To facilitate this, there have been several survey papers, however, none of them covers the aspect of blockchain-cloud integration from a service-oriented perspective. This paper aims to fulfil this gap by providing a service oriented review of blockchain-cloud integration. Indeed, in this survey, we explore different service models into which blockchain has been integrated. For each service model, we review the existing works and present a comparative analysis so as to offer a clear and concise view in each category.

\end{abstract}


\section{Introduction}
Cloud computing has already been a day-to-day purpose technology. It offers on demand services using a pay-as-you-go approach. It gives uninterrupted network access and resource pooling with rapid elasticity \cite{hurwitz2020cloud}. Cloud computing solves the traditional resource management problem by minimizing the cost at an eye-catching rate. However, it still has some limitations such as shared infrastructure problems, virtualization issues, API securities, privacy, SLA (Service Level Agreement) based legal issues and so on \cite{rimal2009taxonomy}. Researchers are trying to solve these problems with the help of different technologies. Recently, Blockchain has emerged as one of the most common technologies in this regard.

Blockchain is regarded as a foundational technology with the potential to disrupt a number of application domains, including cloud computing. It enables a public or private distributed system that holds data in a secure cryptographic fashion \cite{drescher2017blockchain}, thus ensuring a secured transaction mechanism without involving any central entity. Overall, this provides a much better complementary service provision methods for many established service platforms including cloud computing. Therefore, researchers and practitioners are exploring how blockchain and cloud can be integrated to mitigate different issues in cloud environments.

To exploit the full benefits of this cloud-blockchain integration, it is important to have a clear understanding of the impacts blockchain has over different aspects of cloud. There have been a few review papers in this regard which can be found in \cite{han2016secure, park2017blockchain, gai2020blockchain}. Particularly, the review in \cite{gai2020blockchain} is noteworthy as the authors presented a comprehensive survey on different aspects involving blockchain-cloud integration. Unfortunately, a concise service-oriented review of blockchain-cloud integration is missing, even though the majority of the service provision mechanism in cloud relies on a service model. We aim to fulfil this gap in this paper. 

\vspace{1.5mm}
\noindent \textbf{Contributions:} In this survey, we have reviewed a number of works in the cross-section of blockchain and cloud following a service-oriented taxonomy highlighting the service models into which blockchain has been integrated. For each such domain, we have reviewed the existing works and provided a comparative analysis so as to highlight their advantages and limitations. 

\vspace{1mm}
\noindent \textbf{Structure:} The rest of this paper is structured as follows. In Section \ref{sec:back}, we present a brief background of cloud computing and blockchain. Section \ref{sec:tax} presents the service-oriented taxonomy. Afterwards, a review for each service category is presented in Section \ref{sec:saas}, Section \ref{sec:fraas}, Section \ref{sec:faas} and Section \ref{sec:maas}. We discuss and analyze our findings in Section \ref{sec:discussion} and finally, conclude in \ref{sec:conclusion}.
    
\section{Background}
\label{sec:back} 
In this section, we provide a brief background on cloud computing and blockchain.

\vspace{1mm}
\noindent \textbf{Cloud computing:} Cloud services are facilitated by cloud service providers (CSPs) which are basically third parties that provide cloud storage and virtualization as well as other services such as networking components, data, operating systems and so on \cite{hurwitz2020cloud}. CSPs package these services in three fundamental service models: Infrastructure-as-a-Service (IaaS), Platform-as-a-Service (Paas) and Software-as-a-Service (SaaS). Recently other service models such as Blockchain-as-a-service and Security-as-a-service are emerging. There are some useful requirements, such as elasticity, privacy, low-cost, scalability, inter-operability and high performance, that should be kept in mind when designing a cloud architecture. Data management, virtualization management, adaptability and load balancing are also very important factors to consider. 

\vspace{1.5mm}
\noindent \textbf{Blockchain:} Blockchain is a distributed fault-tolerant database where each network participant can share, but no entity can control this \cite{drescher2017blockchain}. It is organized as an append-only list, where every block contains the hash of the previous block, except the first block, called the \textit{Genesis} block. Each of the block encodes some functionalities such as assets and data transfer. Every block is broadcast to the network, verified and added to the existing chain by special nodes or discarded according to the verification result \cite{bashir2017mastering}. The concept of blockchain was popularised with the introduction of Bitcoin \cite{nakamoto2019bitcoin} as an immutable ledger of transactions for a crypto-currency called \textit{Bitcoin}. Since then, Blockchain has evolved from a digital currency to a programmable interactive environment for building distributed reliable applications \cite{tschorsch2016bitcoin}. Some key characteristics which are driving blockchain forward are data immutability (integrity), data provenance, data persistence and distributed consensus \cite{chowdhury2019comparative}.
    
\definecolor{light-gray}{gray}{0.95}
\definecolor{deep-gray}{gray}{0.65}


    

\section{Service-oriented taxonomy}
\label{sec:tax}
\begin{figure}[h]
  \centering
   \includegraphics[trim=12 20 40 15,clip,width=\linewidth]{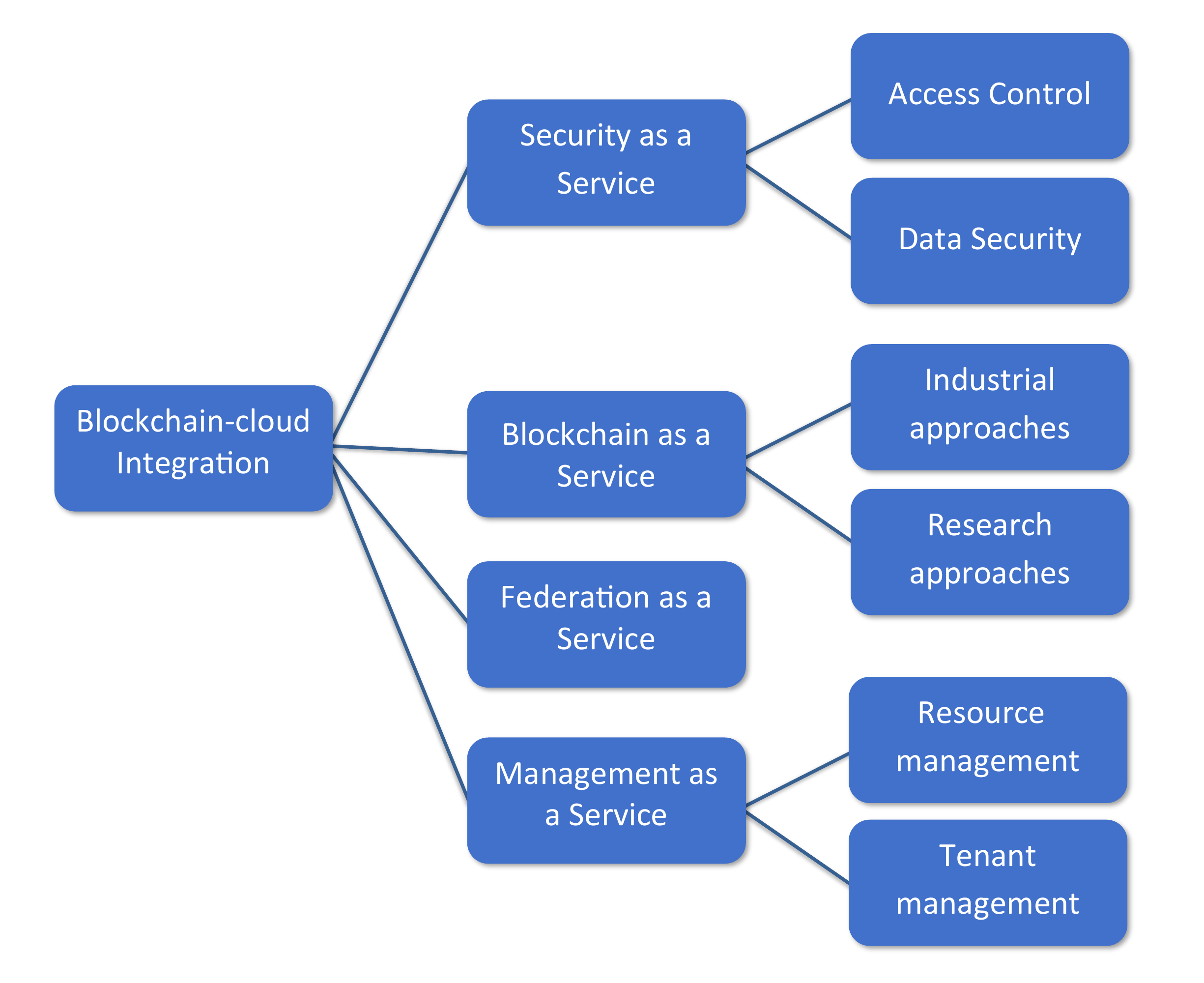}
  \caption{Taxonomy for blockchain-cloud integration}\label{fig:tax}
\end{figure}

This paper focuses on the researches which focus on the blockchain integration with cloud platforms. To review the studies we have created a classification developed under the `as-a-service' perspective, which we call the service-oriented taxonomy. Under this perspective, all researches within the scope of the paper (blockchain-cloud integration) have been grouped together into four service models as illustrated in Figure \ref{fig:tax}. The first group is called \textit{Security as a Service} in which we explore the blockchain-based works that aim to improve the existing security services within a cloud platform. The second group, called \textit{Blockchain as a Service}, deals with those works that offer blockchain services using cloud environments. The third group is a special service model called \textit{Federation as a Service} investigating the works that elaborate on how a blockchain-enabled identify federation can be formed and managed within multiple cloud environments. Our final group consists of research works that focus on the management of tenants and other resources using blockchain within a cloud environment, creating the notion of a \textit{Management as a Service}. For each of these categories, we have reviewed and compared major influential works under different criteria so as to provide a visual analogy of their differences, strengths and weaknesses.

\section{Security as a Service}
\label{sec:saas}
The Security as a Service (SECaaS) model enables the provision of different security services of a cloud platform \cite{varadharajan2014security}. In this section we explore two different types of blockchain-based cloud security services: access control (Section \ref{subsec:userSec}) and data security (Section \ref{subsec:dataSec}).
\subsection{Access control}
\label{subsec:userSec}

Access Control is a fundamental security mechanism used not only for cloud platforms but also for many traditional schemes. It is used to define policies that dictate who can access which data or resources within a system, thereby providing a crucial mechanism to handle unauthorised access in a system. Blockchain-based access control models for cloud aim to tackle two major challenges:
\begin{itemize}
\item In an encryption-based traditional access control model, one trusted central server is required to store access policies governing the rights and permissions obligations as well as to generate, manage and distribute the keys, thereby making it a single point of failure.
\item Securely sharing data/resources using a flexible access control mechanism within a cloud.
\end{itemize}

\setlength{\textfloatsep}{2pt}
\newcolumntype{P}[1]{>{\centering\arraybackslash}p{#1}}

\begin{table*}[!h]
\caption{Comparison of access control mechanisms in blockchain-cloud integration}\label{tab:acControl}
\centering
\rowcolors{1}{deep-gray!25}{white}
\begin{tabular}{P{2cm}|P{1.5cm}|P{1.5cm}|P{1.9cm}|P{1cm}|P{3cm}|P{3.7cm}}
\hline
\rowcolor{deep-gray}
Research               & Technique      & Platform    & Encryp. Method     & FG & Advantages                      & Disadvantages                                                                          \\ \hline \hline
Wang et al. \cite{wang2018blockchain}       & DAC,ABAC       & Ethereum & MIRACL (ABE-80)  & \fillcirc{3pt}  & Feasible, Inter-operable    & No attribute revocation, policy update\\ \hline
Wang et al. \cite{wang2019secure}       & ABAC    & Ethereum   & CPABE      & \fillcirc{3pt}  & No central key distributor & Lack of Integrity\\ \hline
Sukhodolskiy et al\cite{sukhodolskiy2018blockchain} & ABAC     & Ethereum   & ABE       & \empcirc{3pt}  & Dynamic access policy      & Cannot handle a resource with multiple owners\\ \hline
AuthPrivacy \cite{yang2020authprivacychain}        & Identity based & EOS & AES \& Asymmetric & \empcirc{3pt}  & Authorization revocation   & Complex implementation \\ \hline
\vspace{-3mm}
\end{tabular}
\end{table*}

Wang et al. \cite{wang2018blockchain} proposed a blockchain-based framework for data sharing within cloud environments utilising Ethereum (a public blockchain platform, \texttt{https://ethereum.org/en/}), IPFS (a distributed data sharing platform, \texttt{https://ipfs.io/}) and \textit{Attribute Based Encryption }(ABE) mechanism. Their scheme filters un-authorized search requests using an \textit{AND} gate access policy for multiple attribute values and wildcards and allows the data owner to secretly share data in cloud with \textit{public generator keys} (PKG). On a similar note, the authors in \cite{wang2019secure} proposed a cloud storage framework with automated access control enforced using Ethereum and a cryptographic mechanism called \textit{Ciphertext-policy Attribute Based Encryption (CP-ABE)}. This scheme enables a data owner to assign attribute sets and define an access policy with a validity period for a resource by creating and deploying a smart-contract with which the policy is enforced. 

The scheme proposed by Sukhodolskiy et al. \cite{sukhodolskiy2018blockchain} has the ability to define and enforce dynamic access policies using Ethereum within a cloud platform. It utilises a \textit{Certificate Authority} (CA) that generates public and secret keys as a response to a user's requests and an \textit{Attribute Authority} (AA) that is responsible for managing keys for all attributes. One limitation of the scheme is the assumption that each resource has only one owner which is not always true in an organizational setting. AuthPrivacyChain \cite{yang2020authprivacychain} uses EOS blockchain (a public blockchain, \texttt{https://eos.io/}) to store the access control rights. It designs an authentication and authorization revocation process by assuming the blockchain account address as the identity. It can resist different types of external and internal attacks and protect the privacy of users.

\begin{table*}[h!]
    \caption{Comparison of blockchain-based SSE mechanisms in cloud}\label{tab:sse}
    \centering
    \rowcolors{1}{deep-gray!25}{white}
    \begin{tabular}{c|c|c|c|c|c}
    
    \hline
    \rowcolor{deep-gray}
    Research   & Client-side Verification & Server-side Verification & TTP & Updatability & Platform \\ \hline \hline
     Li et al.\cite{li2017searchable} & \fillcirc{3pt} & \empcirc{3pt} & \empcirc{3pt} & \empcirc{3pt} & Bitcoin \\ \hline
    Cai et al.\cite{cai2017towards} & \fillcirc{3pt} & \empcirc{3pt}  & \fillcirc{3pt} & \fillcirc{3pt} & Ethereum\\ \hline
    TKSE\cite{zhang2018tkse} & \fillcirc{3pt} & \fillcirc{3pt} & \empcirc{3pt} & \empcirc{3pt}  & Ethereum and Bitcoin \\ \hline
    
    Zhang et al.\cite{zhang2018outsourcing} & \fillcirc{3pt} & \empcirc{3pt} & \empcirc{3pt} & \empcirc{3pt} & Ethereum and Bitcoin \\ \hline
    Hu et al.\cite{hu2018searching} & - & - & \empcirc{3pt} & \fillcirc{3pt} & Ethereum  \\ \hline
    Chen et al.\cite{chen2019blockchain} & -  & - & \empcirc{3pt} & \fillcirc{3pt} & Ethereum \\ \hline
    \end{tabular}
\vspace{-3mm}
\end{table*}

The summary of our evaluation of the reviewed works under this section is presented in Table \ref{tab:acControl}. In the table, the notations `\fillcirc{3pt}` and `\empcirc{3pt}' have been used to signify if the FG (Fine-grained) property is satisfied or not respectively. Apart from what is presented in the table, there is one issue that is worth mentioning. All these works utilised public blockchain platforms such as Ethereum and EOS. Even though public blockchain platforms provide better security, in comparison to any private one, there is associated cost to store data and smart-contract execution \cite{chowdhury2019comparative}. Furthermore, most public blockchain systems are slow in nature and have scalability issues. Also, public blockchains have privacy issues and all of these works had to separately resolve privacy issues.

\subsection{Data security}
\label{subsec:dataSec}
Schemes under this category aim to ensure the confidentiality, privacy, integrity and provenance of data utilising blockchain within cloud environments.

\vspace{1mm}
\noindent \textbf{Data confidentiality \& privacy:} To maintain the confidentiality and privacy, data must be stored with some pre-processing formats or using any encryption methodology before storing them in cloud. This raises complications while retrieving the encrypted data. In fact, users have to download all data, decrypt them and retrieve the needed portion using a query, which requires additional time and expensive computations for large data size \cite{li2017searchable}. In recent years, SSE \textit{(Secure Searchable Encryption)} schemes have emerged where user’s data are encrypted using a private key and stored in a masked index table with encrypted massages’ keyword pairs. This table and pre-processed data are stored on the server. For retrieving or querying, a search token is generated from the user side and it filters out the encrypted data from the server using that masked index table without needing to decrypt them.

Towards this motivation, Li et al. \cite{li2017searchable} proposed an \textit{``SSE-using-BC''} scheme where the encrypted data is stored in a blockchain-enabled decentralized storage supporting data confidentiality and search efficiency. The data owner can upload encrypted file and its corresponding index to the cloud. The data containing the keyword, will be retrieved in an encrypted manner and the user can decrypt it locally using search tokens. Their proposal can handle both small and large scale data.

Cai at el. \cite{cai2017towards} proposed an encrypted decentralized storage architecture, considering the security and fair payment services utilising a TTP \textit{(Trusted Third Party)}. This scheme, compatible with both Ethereum and Bitcoin, enables the client or the data owner to perform file addition on targeted storage peers with verifiable keyword searching. In TKSE \cite{zhang2018tkse}, they first introduced an SSE scheme with the user and server sided verifiability supporting cost minimization and fair judgement. For the user-sided verifiability, a data owner can integrate search requirements into the output script of joint transactions in such a way that the data owner can resist malicious cloud servers. The server-side verifiability has been achieved using the public verification of digital signature. 

Zhang et al. \cite{zhang2018outsourcing} introduced BPay, a blockchain-enabled outsourcing service framework with payment fairness against malicious users or service providers. Their robust all-or-nothing protocol is based on SSE and is compatible with both Bitcoin and Ethereum. In \cite{hu2018searching}, Hu et al. leveraged an Ethereum smart-contract to replace a central server architecture and constructed an efficient search scheme preserving security with efficient computation. The integrating search algorithm in the smart-contract had focused on two issues: correctness in search result and computational overhead. Following the approach of \cite{hu2018searching}, Chen et al. \cite{chen2019blockchain} focused on data access control of health record data in cloud and presented a blockchain-enabled SSE scheme. They also implemented a smart-contract in Ethereum for fair payment services in a multi-user setting. 

\begin{table*}[h!]
\caption{Comparison of blockchain-based data provenance mechanisms in cloud}\label{tab:prov}
\rowcolors{1}{deep-gray!25}{white}
\begin{tabular}{P{2.8cm}|P{1.2cm}|P{1.8cm}|P{1.2cm}|P{1.3cm}|P{2cm}|P{1.4cm}|P{2.6cm}}
\hline
\rowcolor{deep-gray}
Research  & Scalability &  Customizability & Inter\-operability & Access\-Control & Environment      & Platform   & Monitoring Tool \\ \hline 
Smartprovenance\cite{ramachandran2018smartprovenance} & \fillcirc{3pt}           & \fillcirc{3pt}               & \empcirc{3pt}                & \fillcirc{3pt}             & Google drive     &  Ethereum         & Event listener module \\ \hline
Provchain\cite{liang2017provchain}      & \fillcirc{3pt}           & \fillcirc{3pt}               & \empcirc{3pt}                & \fillcirc{3pt}             & OwnCloud         & Bitcoin         & Hooks API             \\ \hline

BlockCloud\cite{shetty2017data}     & \fillcirc{3pt}           & \fillcirc{3pt}               & \empcirc{3pt}                & \fillcirc{3pt}             & OwnCloud         & PoS blockchain         & Hooks and listeners   \\ \hline
Tosh et al.\cite{tosh2019data}   & \fillcirc{3pt}           & \empcirc{3pt}               & \fillcirc{3pt}                & \fillcirc{3pt}             & Federated cloud & -         & Hooks and listeners   \\ \hline

Zhang et al.\cite{zhang2017blockchain}    & \empcirc{3pt}           & \empcirc{3pt}               & \fillcirc{3pt}                & \empcirc{3pt}             & Cloud forensic   & -           & User                  \\ \hline

Gaetani et al.\cite{gaetani2017blockchain} & \fillcirc{3pt}           & \empcirc{3pt}               & \fillcirc{3pt}                & \fillcirc{3pt}             & Distributed DB   & - & Database              \\ \hline

CPVPA\cite{zhang2019blockchain}          & \empcirc{3pt}           & \fillcirc{3pt}               & \empcirc{3pt}                & \fillcirc{3pt}             & Public cloud & -           & Auditor               \\ \hline
\end{tabular}
\vspace{-3mm}
\end{table*}

The summary of blockchain-supported SSE based research works is presented in Table \ref{tab:sse}. Like before, the notations `\fillcirc{3pt}' and `\empcirc{3pt}' have been used to signify if a property is satisfied or not respectively whereas `-' implies not applicable. As per the table, only Cai et al. utilised a TTP and all are based on public blockchain platforms (mostly Ethereum). Also, when applicable, a client-side verification issue has been addressed whereas the server-side verification has been considered in only TKSE scheme.

\vspace{1mm}
\noindent \textbf{Data integrity and provenance:} Data integrity is the maintenance and assurance of the accuracy, completeness and consistency of data whereas data provenance generally describes the custodial chronology of an object by recording every information of a data object, from its creation to modification to deletion and then storing this information in a verifiable audit trail. The core idea of utilising blockchain for these two is to use the immutability and transparency properties of blockchain to record each activities associated with a data.

Smartprovenance \cite{ramachandran2018smartprovenance} uses an \textit{open provenance method} (OPM) and introduces a fully automated verification scheme using two smart-contracts. The \textit{document tracker} contract, as its name says, keeps track of the changes of a particular document while the \textit{voting contract} is used to implement \& initiate the voting protocols. The proposal was deployed using Ethereum and Google Drive cloud storage.   

Provchain \cite{liang2017provchain} offers a solution for collecting, storing and verifying provenance data in a highly scalable way in which auditors are used for the verification process and for answering the provenance-related queries. It was implemented using Bitcoin and \textit{OwnCloud} (a cloud based file sharing service, \texttt{https://owncloud.com/}). BlockCloud \cite{shetty2017data} architecture is almost similar to Provchain, except that it utilised a \textit{Proof-of-Stake} (PoS) consensus algorithm for reducing high electricity usage in \textit{Proof-of-WorK} (PoW) consensus algorithm. A centralized entity called \textit{Federation service} controls the resources and manages the process of stake determination, allocation \& verification. 

Tosh et al. \cite{tosh2019data} proposed an integrity checking mechanism with provenance data by querying searchable provenance database through the auditor within a cloud. Zhang et al. \cite{zhang2017blockchain} proposed a process provenance, which provides the proof of existence and privacy preservation for process records using blockchain and group signature.

A two-layered blockchain within a cloud environment was proposed by Gaetani et al. \cite{gaetani2017blockchain}, where the first layer was a mining rotation-based blockchain to improve the performance whereas the second layer was a PoW-based as-is blockchain to ensure the integrity. CPVPA \cite{zhang2019blockchain} introduced a certificate-less public veriﬁcation scheme against procrastinating auditors. Their scheme removes the trusted third-party auditor in data integrity checking, instead introduces a dedicated trust-less third-party auditor which can itself be audited easily (using block generation time) by any user. Because of its certificate-less mechanism, it avoids the certificate management problem as well.  

A comparative summary of the reviewed works under this category is presented in Table \ref{tab:prov} where the notations have their usual semantics. 


\setlength{\textfloatsep}{2pt}
    \begin{table*}[h!]
    \centering
    \caption{Comparison of industrial BaaS applications}\label{tab:baas}
    \rowcolors{1}{deep-gray!25}{white}
    \begin{tabular}{c|c|c|c}
    \hline
    \rowcolor{deep-gray}
    Services & Platform   & Type   & Scalability    \\ \hline \hline
    Azure\cite{azure_ref}        & Ethereum, Corda, Fabric     & Private   & Auto scaling on service demand  \\   \hline
    IBM\cite{ibm_ref}        & Fabric     & Private   & High scalability with IBM cloud  \\   \hline
    AWS\cite{aws_ref}        & Ethereum, Corda, Fabric    & Private   & Quick node creation APIs      \\   \hline
    HPE\cite{hpe_ref}        & Corda     & Private   & Linear scalability with backup facilities    \\   \hline
    Oracle\cite{oracle_ref}       & Fabric     & Private, Consortium   & Dynamic scaling for all resources    \\   \hline
    SAP\cite{sap_ref}        & Fabric     & Private   & High scalability with SAP HANA facilities   \\   \hline
    Google\cite{google_ref}        & Ethereum, Fabric     & Private   & Scaling with Google App Engine   \\   \hline
    
    \end{tabular}
    \vspace{-3mm}
    \end{table*}

\section{Blockchain as a Service}
\label{sec:fraas}
 Blockchain as a Service (BaaS) can be considered as a service like SaaS, however, instead of providing a particular software service, BaaS enables the customers to create, deploy and maintain blockchain networks within a cloud environment. With the increasing adoption of blockchain in business plans of different sectors, the CSPs have seen the benefits of providing blockchain based services to their users. Researchers have also started exploring how such services can be improved. We segregate and explore these industrial (Section \ref{subsec:research}) and research (Section \ref{subsec:research}) schemes next. 

\subsection{Industrial Approaches}
\label{subsec:industry}
Almost all major CSPs such as Microsoft, Amazon and IBM are providing BaaS. \textit{Microsoft Azure} \cite{azure_ref} is considered as a pioneer in this domain when they introduced EBaaS (Ethereum BaaS). Now Azure also supports other private blockchain platforms such as Hyperledger Fabric (\texttt{https://www.hyperledger.org/use/fabric}) and Corda (\texttt{https://www.corda.net/}). \textit{IBM Blockchain} \cite{ibm_ref} is rendering BaaS using Hyperledger Fabric. \textit{Amazon AWS} \cite{aws_ref} also offers BaaS in two forms: i) \textit{Amazon Quantum Ledger Database} (QLDB) which enables users to create customised blockchains and ii) \textit{Amazon Managed Blockchain} (AMB) using Hyperledger Fabric. Hewlett Packard Enterprise also introduced a \textit{``Mission Critical Distributed Ledger Technology''} \cite{hpe_ref} to provide BaaS backed by Corda. Similarly, Oracle \cite{oracle_ref}, SAP \cite{sap_ref} and Google \cite{google_ref} have also launched their Baas offering where Oracle and SAP are utilising Hyperledger Fabric whereas Google is using both Hyperledger Fabric and Enterprise Ethereum (a private blockchain initiative). The summary of different BaaS is presented in Table \ref{tab:baas}.

\setlength{\textfloatsep}{2pt}
\begin{table*}[h!]
\centering
\rowcolors{1}{deep-gray!25}{white}
\caption{Comparison of FaaS applications}\label{tab:faas}
\begin{tabular}{c|c|c|c|c}
\hline
\rowcolor{deep-gray}
Research             & Platform    & Purpose                                    & Inter-operability & Privacy \\
\hline \hline
Schiavo et al.\cite{schiavo2016faas} & Any & Democratic governance on data and services & \empcirc{3pt}          & \fillcirc{3pt}       \\
Alansari et al.\cite{alansari2017distributed} & Ethereum    & Finegrained attribute-based access control & \empcirc{3pt}          & \fillcirc{3pt}       \\
Alansari et al. \cite{alansari2017privacy}     & Ethereum    & Attribute based access control             & \empcirc{3pt}          & \fillcirc{3pt}       \\
Yang et al. \cite{yang2018differentially}     & Fabric & Differntially-private data sharing        & \fillcirc{3pt}          & \fillcirc{3pt}       \\
DRAMS \cite{ferdous2017decentralised}           & Ethereum    & Decentralised runtime monitoring           & -          & \fillcirc{3pt}  \\

Margheri et al.\cite{margheri2017distributed}  & Ethereum    & Governance for democratic control           & -          & \fillcirc{3pt}       \\
CloudChain \cite{ghosh2019towards}      & Fabric & Democratic control on IaaS provisioning    & \fillcirc{3pt}          & \empcirc{3pt}       \\

\hline
\end{tabular}
\vspace{-3mm}
\end{table*}

\subsection{Research Approaches}
\label{subsec:research}
Chen et al. \cite{chen2018fbaas} proposed \textit{Functional BaaS} (FBaaS) on serverless architecture focusing on storage overhead and business logic stages. Apart from other conventional BaaS models, \textit{Big Data Open Architecture} (BDOA) with four layers has been incorporated with this model. Chen et al. \cite{chen2019full} presented \textit{Full-Spectrum Blockchain as a Service} (FSBaaS) which is a platform to enhance both centralized and decentralized business collaborations using blockchain technologies. NutbaaS \cite{zheng2019nutbaas} is another similar platform providing services, such as network deployment, system monitoring, smart contact analysis and testing, within a more reliable and secure environment. Their four-layer model supports Hyperledger Fabric, Ethereum, EOS and Filecon and thus, enabling users to combine different blockchain types. uBaaS \cite{lu2019ubaas} is a \textit{unified BaaS} platform which aims to address the issue of vendor-lockins within the existing BaaS models. uBaaS facilitates an independent model including a one-click deployment as a service, design pattern as a service and auxiliary services. These works have similar features and differ only in the number of supported blockchains.

\section{Federation as a service}
\label{sec:faas}
Cloud federation refers to the unification of different services from different providers who may come from disparate networks. It gives the users a more flexible service-delivery option with enhanced availability. In a federated cloud, organizations share resources across their infrastructures and and users can access those services via the internet. The notion of \textit{Federation as a Service} (FaaS) was first introduced in \cite{schiavo2016faas} in which the authors presented a blockchain-enabled decentralised federated governance model. This model emphasizes on the role of blockchain for the secure, transparent and accountable creation and management of cloud federations as well as data and services in a cloud federation.

Other researchers explored how blockchain could be leveraged for different aspects of cloud federations. For example, Alansari et al. presented an identity and access management system for secure data sharing within a cloud federation \cite{alansari2017distributed}. Their approach adopted \textit{Intel's SGX}, a trusted hardware environment, for protecting the integrity and confidentiality of the policy enforcement process. In another work, Alansari et al. \cite{alansari2017privacy} proposed a privacy preserving access control framework for a cloud federation which utilises blockchain, \textit{Oblivious Commitment Based Envelope} (OCBE) protocol and a two-phase \textit{Pedersen commitment} scheme for ensuring security and privacy. Yang et al. \cite{yang2018differentially} proposed a blockchain-based privacy preserving data sharing mechanism within a cloud federation. Their proposal utilises differential privacy to anonymize data before sharing where a smart-contract in the blockchain platform is used to validate and allocate privacy budgets in order to balance between the utility and privacy of the data.  

DRAMS\cite{ferdous2017decentralised} proposes a blockchain-based decentralized runtime access monitoring system for a federated cloud in order to ensure that the components that receive, process and exchange access requests can not be subverted. Margheri et al. \cite{margheri2017distributed} proposed an innovative governance approach, data masking, anonymization and access control monitoring services, to ensure a democratic control of different providers in a cloud federation. On the other hand, CloudChain\cite{ghosh2019towards} proposed a blockchain-based democratic infrastructure service provisioning system for a cloud federation ensuring the transparency and immutability in resource and information exchange.

    
The reviewed works are summarized in Table \ref{tab:faas} where the symbols carry the usual semantics. As evident from the table, most of approaches have strong support for privacy and utilised Ethereum, however, two have adopted Hyperledger Fabric for their deployment.

\setlength{\textfloatsep}{2pt}
\begin{table*}[h!]
    \caption{Comparison of blockchain-based resource and tenant management in cloud}\label{tab:maas}
    \centering
    \rowcolors{1}{deep-gray!25}{white}
    \resizebox{\textwidth}{!}{
    \begin{tabular}{c|c|c|c}

    \hline
    \rowcolor{deep-gray}
    Research   & Scope & Services & Smart-contract Dependency  \\ \hline \hline
    Xu et al. \cite{xu2017intelligent}                                  & Cloud datacenters    & Energy cost minimization & Decision making on power utilization             \\ \hline
    Xiong et al.\cite{xiong2018cloud}   & Local Computaional resources    & Profit maximization and utility management       & Resource auction and mining tasks           \\ \hline
    Nayak et al. \cite{nayak2018saranyu}        & Users' service management    & Service and pricing management & Services and pricing monitoring              \\ \hline
    Weber et al. \cite{weber2019platform}         & Users' privacy  & Individual permissioned blockchain           & Managing private and public blockchain             \\ \hline
    
    \end{tabular}
    }
\vspace{-3mm}    
\end{table*}

\section{Management as a service}
\label{sec:maas}

In a broader sense, the management in cloud can be referred to the administrative controls which encapsulate the process of evaluating and monitoring of users, data, resources, applications, and services. In this section, we review those research works which focus on the utilization of blockchain for different aspects of cloud management. 

\vspace{1.5mm}
\noindent \textit{Resource Management} on cloud comprises of energy optimization, capacity allocation and load balancing between several distributed nodes. Xu et al. \cite{xu2017intelligent} proposed a blockchain-based decentralized resource management framework to minimize the total cost of energy consumption within a cloud. The main feature of this framework is that its consensus process is not similar to PoW or PoS, instead, it allows all participants in the cloud to participate in the consensus process. 
Xiong et al. \cite{xiong2018cloud} on the other hand proposed a game theoretic approach for achieving efficiency over computational resource management. To achieve the desired optimization, they implemented two staged \textit{Stackelberg game} for the consensus process. This unique game theoretic approach enables the cloud provider to set reasonable pricing over computational resources.

\vspace{1mm}\noindent \textit{Tenant Management} is an important issue in cloud for both single or multi-tenancy environments. Nayak et al. \cite{nayak2018saranyu} claimed their system, \textit{Saranyu}, to be the first smart contract based application on account management for cloud tenants. Weber et al. \cite{weber2019platform} proposed an architecture for multi-tenant blockchain-based system where every tenant is assigned an individual permissioned blockchain network to maintain the privacy for their data and smart-contract. 

The reviewed works in this category are summarised in Table \ref{tab:maas} focusing on their scopes, services and smart-contract utilisation.

\section{Discussion}
\label{sec:discussion}
In this survey, we have reviewed a number of influential works in the cross-section of blockchain and cloud integration using a service-oriented taxonomy. Next, we highlight some of our observations that we have noted while carrying out this survey.

\vspace{1mm}
\noindent \textbf{Security:} The majority of the research works in blockchain-cloud integration focus on the security aspects where blockchain has been used to satisfy, mitigate or improve one or more security issues. This is not surprising because of the number of security advantages offered by blockchain in the form of immutability (integrity), provenance, no single point of failure and so on.

\vspace{1mm}
\noindent \textbf{Blockchain platforms:} As mentioned in Section \ref{subsec:userSec}, public blockchains have scalability and privacy issues and a significant cost is incurred for computation and data storage in such platforms \cite{chowdhury2019comparative}. Despite these, a surprising observation from our review is that the majority of the deployments have utilised public blockchain platforms. A possible reason could be that a stable private blockchain platform such as Hyperledger Fabric was not available two/three years ago, forcing the researchers to experiment with a public blockchain platform. However, we expect a change in this regard in future as more and more private blockchain platforms emerge in the market.

\begin{figure}[h]
\centering
\includegraphics[trim=15 0 15 30, clip, width=0.45\textwidth]{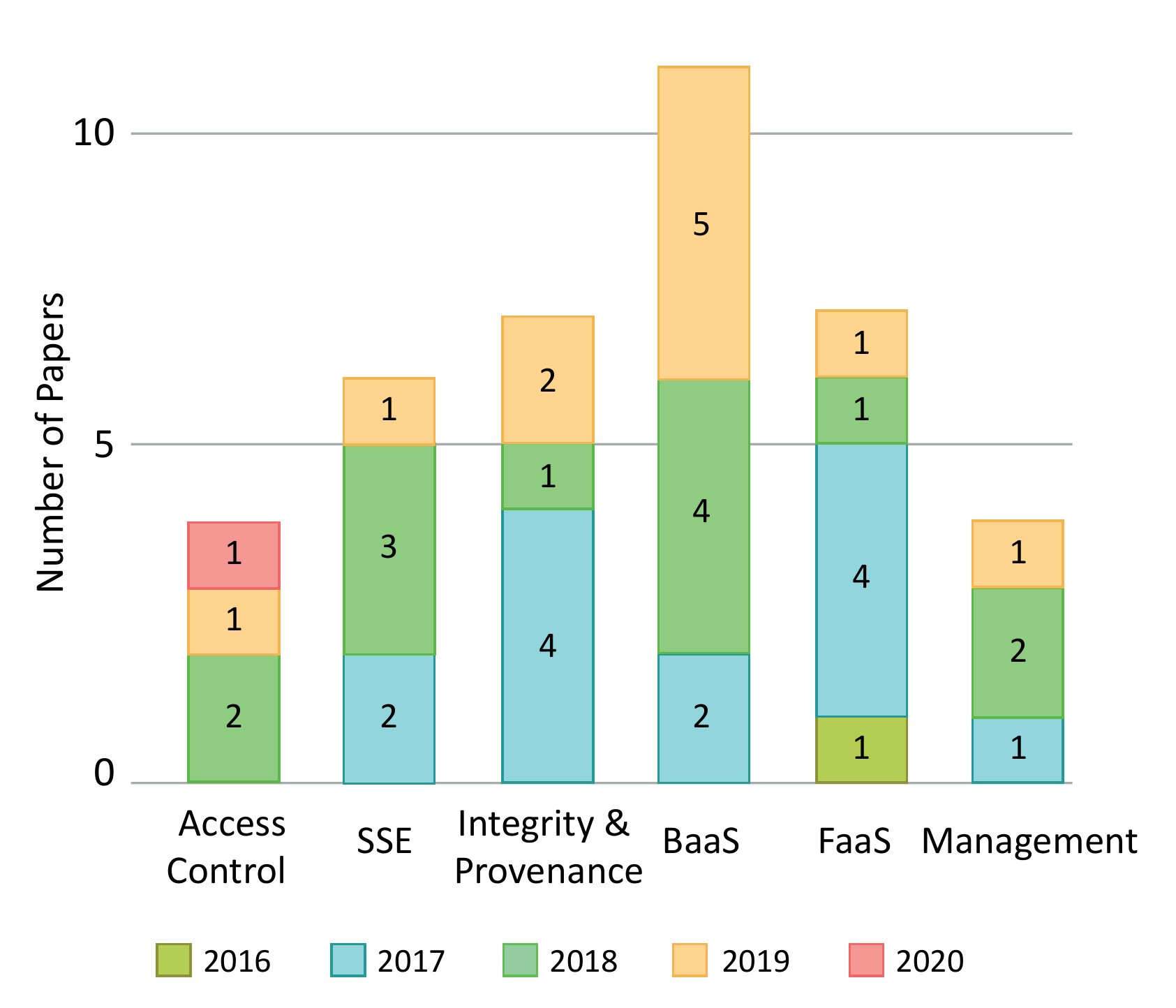}
\caption{Year-wise distributions of blockchain-cloud integration research}\label{fig:year}
\vspace{-2mm}
\end{figure}

\vspace{1.5mm}
\noindent \textbf{Trend analysis:} The year-wise distribution of research (Figure \ref{fig:year}) illustrates interesting trends and provides an indication of the time-frame of the works within a service model. 

\vspace{1.5mm}
\noindent \textbf{Novel service models:} Blockchain has not only been used to mitigate existing issues in cloud. Indeed, it has facilitated the creation of novel service models such as BaaS or even FaaS where blockchain remains at the heart of service delivery model. We expect to see more such novel cloud service models, underpinned by blockchain, in future.

\section{Conclusion}
\label{sec:conclusion}
With the increasing growth in the use of cloud platform and the popularity of blockchain technology, different CSPs and researchers are exploring the ways to integrate these two technologies. In our survey, we have reviewed a number of influential works using a service oriented taxonomy in the cross-section of these two technologies. Under each category of the taxonomy, we have summarized the reviewed works in a table so as to provide a side by side comparison of the works against a set of properties. Such comparison between existing approaches within a service model will help the practitioners and researches to analyze the paradigms in an efficient way and thereby, providing a concise understanding of the research gaps which will ultimately help them to explore exciting researches in this domain.

\bibliographystyle{IEEEtran}
\bibliography{main}

\end{document}